\documentclass{JINST}

\title{Sub-millimeter nuclear medical imaging with high sensitivity
       in positron emission tomography using $\beta^+\gamma$ coincidences}
\author{C. Lang$^a$, D. Habs$^{a,b}$, K. Parodi$^a$
        and P.G. Thirolf$^a$\thanks{Corresponding author}\\
\llap{$^a$}Fakult\"at f\"ur Physik (LS f. Medizinische Physik),
           Ludwig-Maximilians Universit\"at M\"unchen,
           Am Coulombwall 1, D-85748 Garching, Germany\\
\llap{$^b$}Max-Planck-Institut f\"ur Quantenoptik,
           Hans-Kopfermann-Str. 1, D-84748 Garching, Germany\\
  E-mail: \email{Christian.Lang@physik.uni-muenchen.de}}

\abstract{
We present a nuclear medical imaging technique,
employing triple-$\gamma$ trajectory intersections from $\beta^+ - \gamma$ coincidences,
able to reach sub-millimeter spatial
resolution in 3 dimensions with a reduced requirement of reconstructed intersections
per voxel compared to a conventional PET reconstruction analysis.
This '$\gamma$-PET' technique draws on specific $\beta^+$-decaying isotopes,
simultaneously emitting an additional photon. Exploiting
the triple coincidence
between the positron annihilation and the third photon, it is possible to separate the
reconstructed 'true' events from background. In order to characterize this
technique, Monte-Carlo simulations and image reconstructions have been performed. The
achievable spatial resolution has been found to reach ca.~0.4 mm (FWHM) in each direction
for the visualization of a $^{22}$Na point source.
Only 40 intersections are sufficient for a reliable sub-millimeter image
reconstruction of a point source embedded in a scattering volume of water
inside a voxel volume of about 1~mm$^3$ ('high-resolution mode').
Moreover, starting with an injected activity of 400 MBq for $^{76}$Br,
the same number of only about 40 reconstructed intersections are needed in case of
a larger voxel volume of 2~x~2~x~3~mm$^3$ ('high-sensitivity mode').
Requiring such a low number of reconstructed events significantly reduces the
required acquisition time for image reconstruction (in the above case to about 140~s)
and thus may open up the perspective for a quasi real-time imaging.}


\begin{document}

\section{Introduction}

Functional medical imaging via positron emission tomography (PET) experienced an
enormous growth in the last decade due to the large variety of
different tracers and due to the significant improvement of the imaging
performance~\cite{Suetens02}.
These tracers are chemical compounds, carrying a positron emitting isotope to, e.g., a
tumor site. Annihilation of the positron into two back-to-back 511 keV photons allows
to restrict the source origin in 2 dimensions onto a line of response (LOR). 
Superimposing
the LORs of different decay events locates the source distribution of the emitter, i.e.
the concentration of the tracer molecules in a patient in 3D. 
Modern whole-body systems are combined PET/CT scanners, where CT provides the anatomical
information for a reference frame and also data for the attenuation and scattering
correction, while PET provides the molecular information.
Such scanners are able to reach a spatial resolution of about 5-6 mm 
\cite{philips-tf,ge-discovery}.
The performance of new generation PET/CT 
with point-spread-function (PSF) correction can go down to 2.0 mm for the PET 
modality~\cite{siemens-mCT}.
An improvement of conventional PET imaging is the Time-of-Flight (TOF) PET technique, where the
flight time difference of the two annihilation photons registered in the responding
pair of detectors is taken into account for achieving an improved spatial information.
Annihilation events can be restricted to a certain area of the LOR, thus achieving
an improved image quality by an improved signal-to-noise ratio. Also the patient
dose and the examination time can be reduced.
Moreover, especially biomedical research, using animal models of metabolism and disease
mechanisms, would profit from a high-resolution small-animal PET device,
where presently a resolution of 1.1~mm
(FWHM) in the center of the field-of-view (FOV) has been reported, which degrades 
to 2.2 mm (FWHM) at a position with 2 cm radial offset~\cite{Park07, Yang04}. 
Even sub-millimeter spatial resolution may be reachable, as, e.g., first tests with
continuous LYSO crystals 
coupled to a silicon photomultiplier matrix
have shown~\cite{llosa-mic12}.
Also the concept of a Compton camera has been investigated for its applicability with the PET
technique, however, so far only at the laboratory development level~\cite{Grignon07}.
In this case, position and energy information of the Compton scattering kinematics are
measured in a scatter and an absorber detector,
together allowing for a reconstruction of the direction to the source position on the
surface of the 'Compton cone'. 
In general, the spatial resolution of a PET device is limited by several effects. Limiting
factors are the spatial, temporal or energy resolution of the
detectors, random coincidences or the resolving power of the source image reconstruction
algorithm.
Moreover, there are also inherent physical limits to the achievable image resolution, like
Compton scattering of the 511 keV photons within the patient or biological sample or the
diffusion range of the positron before its annihilation,
presenting the dominant limitation in small-animal PET imaging.
Another inherently limiting factor is the acollinearity of the
positron annihilation, i.e. the angular deviation from 180$^o$ between the two
annihilation photons, originating from the
momentum distribution of the annihilating electron-positron pair, after thermalization
 of the positron (within a few ps) and positronium formation.
Because of the thermalization of the positron prior to its annihilation, the acollinearity
is mainly caused by the significantly higher momentum of the bound orbital electrons.
In a recent study~\cite{Shibuya2007} the acollinearity describes the angular deviation
from 180$^o$ by two components: a broad main component originating from orbital electrons with
$\Delta\theta=$0.633(8)$^o$ and a narrower component with $\Delta\theta=$0.27(10)$^o$)
resulting from positronium annihilation. 
This corresponds to a spatial deviation of 2 mm in an average PET-ring radius of 40 cm
and thus represents the dominant limiting
factor for the spatial resolution of whole-body PET systems~\cite{Oxley09}.
Another effect arising from the momentum distribution of the annihilating electron-positron
pair is the Doppler broadening of the annihilation spectrum. The actual broadening depends on the
annihilation medium and can be used as a measure for the longitudinal component of the
momentum distribution~\cite{Torrisi97, Iwata97}.
Besides the achievable spatial source reconstruction resolution, also the applied radioactive
dose to the patient or sample, as well as the corresponding examination time, has to be taken
into account when discussing medical imaging techniques.
The applied radioactivity typically used in human PET studies is tracer specific and ranges
from about 185 to 1850 MBq~\cite{Hutchins00}, while in small-animal PET even higher
activities are applied.
 A typical human PET examination using the radioisotope $^{18}$F takes 10-45 minutes
depending on scanner, tumor and image reconstruction method.
While on the one hand such long examination times limit the number of patients' access to
PET devices, they are prohibitive for real-time metabolism studies, due to, e.g., organ
movements.
Therefore, this study was motivated not only by aiming at an improved spatial resolution for
PET examinations, but, perhaps even more attractive, by targeting
a higher sensitivity via a lower reconstruction statistics required
per voxel of the examination volume.
Moreover, the technique described in the following also bears the potential to
be applicable for ion beam range monitoring in
hadron therapy~\cite{Enghardt04, parodi07b}, exploiting the
online generation of $\beta^+$ emitting isotopes.

\section{Decay properties of PET isotopes}

Table~\ref{tab:Isotope-Range} compares the decay properties of various presently used or
potential future PET radioisotopes.
The two isotopes $^{10}$C and $^{14}$O have been included here, despite of their
short halflives of 19.3~s and 70.6~s, respectively, since both can be produced during
hadron therapeutic irradiations using a carbon
or proton beam~\cite{sommerer09,parodi07}. Thus they qualify as candidates for
online ion-beam range monitoring during therapy treatment.
Moreover, proposals have been presented to directly use positron emitter beams, such as
$^{11}$C~\cite{urakabe01,kanazawa02,tomitani03}, $^{10}$C~\cite{mizuno03} or $^{15}$O
~\cite{inaniwa05} as therapeutic beams, allowing for fast online ion-beam range verification.
In Tab.~\ref{tab:Isotope-Range} in particular the positron diffusion range in water has
been simulated with Geant4 (last column) and benchmarked against measured values
(column 6), showing good agreement.
\begin{table}[h]
\begin{center}
\begin{tabular}{llccccc}
Isotope & decay & $E_{e^+}^{max}$ & I$_{\beta}$ & $E_{\gamma}$ & mean range & mean range\\
        & mode  &                 &              &              &   in water & in water \\
        &       & [MeV]           &  [$\%$]      &  [MeV]       &  [mm]      & [mm]     \\
        &       & \cite{ENSDF02}  &\cite{ENSDF02}&\cite{ENSDF02}&(experiment,&(simulation,\\
        &       &                 &              &              &\cite{Humm03})& this work) \\
\hline
$^{22}$Na & $\beta^{+} + \gamma$ & 0.54      & 100       & 1.27 &       &     1.5 $\pm$ 0.1 \\
$^{18}$F  & $\beta^{+} $         & 0.63      & 96.7      &      & 1.4   &     1.4 $\pm$ 0.1 \\
$^{94}$Tc & $\beta^{+} + \gamma$ & 0.81/1.83 & 10.5/70.8 & 0.87 &       &     1.4 $\pm$ 0.1 \\
$^{11}$C  & $\beta^{+} $         & 0.96      & 99.8      &      & 1.7   &     1.8 $\pm$ 0.1 \\
$^{13}$N  & $\beta^{+} $         & 1.20      & 100       &      & 2.0   &     1.9 $\pm$ 0.1 \\
$^{44}$Sc & $\beta^{+} + \gamma$ & 1.47      & 94.3      & 1.16 &       &     2.1 $\pm$ 0.1 \\
$^{15}$O  & $\beta^{+} $         & 1.73      & 99.9      &      & 2.7   &     2.6 $\pm$ 0.1 \\
$^{14}$O  & $\beta^{+} + \gamma$ & 1.81      & 99.2      & 2.31 &       &     2.6 $\pm$ 0.1 \\
$^{68}$Ga & $\beta^{+} + \gamma$ & 1.90      & 88.0      & 1.08 &       &     2.7 $\pm$ 0.1 \\
$^{124}$I & $\beta^{+} + \gamma$ & 1.53/2.14 & 11.7/10.8 & 0.60 &       &     2.9 $\pm$ 0.1 \\
$^{10}$C  & $\beta^{+} + \gamma$ & 2.93      & 98.5      & 0.72 &       &     2.6 $\pm$ 0.1 \\
$^{152}$Tb& $\beta^{+} + \gamma$ & 2.62/2.97 &  5.5/6.2  & 0.34 &       &     3.6 $\pm$ 0.1 \\
$^{86}$Y  & $\beta^{+} + \gamma$ & 1.22/1.55/& 11.9/5.6  & 1.08 &       &     2.3 $\pm$ 0.1 \\
          &                      & 1.99/3.14 & 3.6/2.0   &      &       &               \\
$^{76}$Br & $\beta^{+} + \gamma$ & 0.87/0.99 & 6.3/5.2   & 0.56 &       &     4.1 $\pm$ 0.1 \\
          &                      & 3.38/3.94 & 25.8/6.0  &      &       &               \\
$^{82}$Rb & $\beta^{+} + \gamma$ & 4.39      & 100       & 0.78 &       &     4.9 $\pm$ 0.1 \\
\end{tabular}
\caption{Decay properties of presently used or potential future PET isotopes. The positron
         diffusion range has been simulated with Geant4 (last column) and compared to
         experimentally measured values, where available.}
\label{tab:Isotope-Range}
\end{center}
\end{table}
We used Geant4 (9.4) with the QGSP-BIC-HP physics list for hadronic interactions and the
Livermore physics list for electromagnetic interactions~\cite{Agnostinelli03}.\\
In case of the positron range simulations, we placed the detector as close as possible to a
water sphere of 6 cm diameter, thus ensuring to be in the regime where diffusion of the
positron prior to its annihilation is the dominant factor for the position resolution, like
in small-animal PET devices. $^{22}$Na is the only non-medical radioisotope listed in
Tab.~\ref{tab:Isotope-Range} due to its use in our laboratory as test source
for the later-on discussed $\gamma$-PET technique.

\section{The $\gamma$-PET imaging technique}
 \label{sec:gpet}

The novel '$\gamma$-PET' imaging technique presented here draws on specific e$^{+}$ sources, 
simultaneously emitting
an additional photon with the $\beta^{+}$ decay. Exploiting the triple coincidence between
the positron annihilation and the additionally emitted photon, it is possible to efficiently
separate the reconstructed 'true' events from background~\cite{Grignon07,Habs11}.
Therefore, the image reconstruction sensitivity can be significantly increased by an improved
signal-to-noise ratio, achieved via
exploiting the spatial and temporal coincidence with the additionally emitted
photon. 

\begin{figure}
  \centerline{\includegraphics[width=0.9\textwidth]{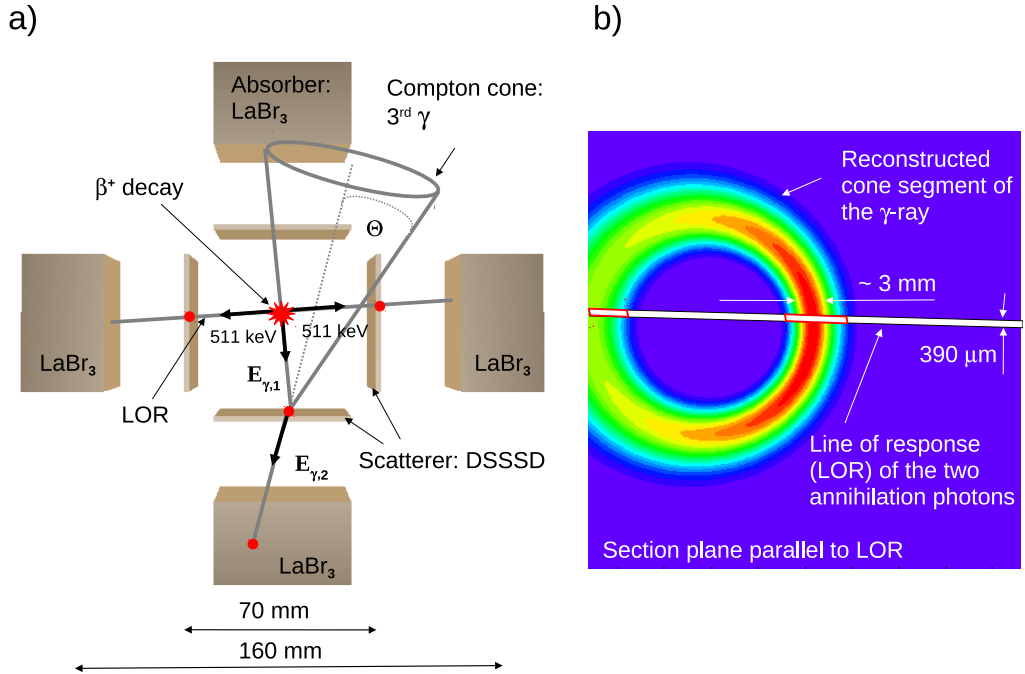}}
  \caption{Principle of the $\gamma$-PET technique (for details see text).
     }
\label{fig:principle}
\end{figure}

Radioisotopes like $^{94(m)}$Tc, $^{67}$Br, $^{124}$I, $^{86}$Y, $^{152}$Tb,
$^{52}$Mn, $^{82}$Rb and $^{44}$Sc are suitable candidates for the $\gamma$-PET technique (see
Tab.~\ref{tab:Isotope-Range} for details).
Especially $^{44}$Sc is of interest, which $\beta^{+}$-decays into the stable $^{44}$Ca,
emitting an 1157 keV photon. It has already been tested clinically~\cite{Roesch11}. With a
short half-life of 3.9 h it has to be produced from a $^{44}$Ti generator
(t$_{1/2}=$60.4a)~\cite{Pruszynski10}, which presently cannot be performed in clinically
relevant quantities. However, this may change
with the soon expected availability of highly brilliant $\gamma$ beams~\cite{Habs11}.
Due to the kinematics of the Compton scattering process and subsequent photon absorption, a
Compton camera allows for reconstructing the origin of a primary photon on the surface of the
'Compton cone'. Superimposing different cones from different events reduces the reconstructed
source distribution in 3 dimensions to the few-millimeter range~\cite{Kanbach05}.
The $\gamma$-PET technique is different, as it will intersect the Compton cone with the 
line of response (LOR) from the same $\beta^{+}$ annihilation $\gamma$ coincidence event, 
thus allowing to reconstruct the source distribution in 3 dimensions from individual events.
The principle of the $\gamma$-PET technique can be seen in the left panel of
Fig.~\ref{fig:principle}. 
The emitted $\gamma$ ray will first be Compton-scattered in a position-sensitive double-sided
silicon strip detector (DSSSD).
Subsequently, the $\gamma$ ray will be absorbed in a position-sensitive LaBr$_3$ scintillator, 
again measuring position and energy of this final interaction. 
The e$^{+}$ annihilation into two (almost) back-to-back 511 keV photons defines the LOR. 
The intersection of the Compton cone with the LOR restricts the source origin in 3
dimensions within one $\beta^{+}\gamma$ coincidence event, as shown in
Fig.~\ref{fig:principle}b).
Figure~\ref{fig:principle}a) displays the schematical geometry of a $\gamma$-PET setup as
used in our simulations, consisting of four Compton cameras, 
placed around a $\beta^{+}$ source isotope emitting a positron and a prompt $\gamma$ ray.
The energy loss $\Delta E_{\gamma,1}$, the residual energy E$_{\gamma,2}$ and the
interaction positions of the Compton scattering process of the prompt $\gamma$ are measured
in a double-sided silicon strip detector
(DSSSD) and an absorbing scintillator (LaBr$_{3}$), respectively.
The initial photon energy E$_{\gamma,1}$ can be calculated from summing $\Delta E_{\gamma,1}$
and E$_{\gamma,2}$.
The opening angle $\theta$ of the Compton-scattering cone can be derived 
according to
\begin{equation}
  \cos{\theta}=1-m_{e}c^{2}\frac{E_{\gamma,2}}{E_{\gamma,1}(E_{\gamma,1}-E_{\gamma,2})}.
\end{equation}
This formula (in contrast to the MC simulations)
is not taking into account the effect of Doppler broadening, arising from
the non-zero momentum of the bound electron.
The Doppler broadening contributes
to the physical limits of the achievable angular resolution
of a Compton camera~\cite{Zoglauer06}.
In case of an unknown initial photon energy E$_{\gamma,1}$, which could occur if the
$\beta^+\gamma$-emitters were produced during hadron therapy with, e.g., a $^{12}$C beam
(potentially generating different $\beta^+\gamma$-emitters like $^{10}$C or $^{14}$O),
proper event and image reconstruction would require a full absorption of E$_{\gamma,2}$
in the scintillator.
The intersection (Fig.~\ref{fig:principle}b) of the Compton cone and the LOR strongly
suppresses background and restricts the reconstructed events to those belonging to
the same $\beta^{+}\gamma$ coincidence event, originating from a volume defined by the
displacement between the positions of the $\beta^+$ decay and the
positron annihilation, depending on the time resolution of the detector system.
In contrast to the restriction of the photon emission volume, the acollinearity effect
 cannot be removed by the $\gamma$-PET technique.
Also Compton scattering of the 511 keV photons within the patient limits the performance.
A further improvement of the $\gamma$-PET technique (so far not implemented in our
reconstruction code) would take attenuation and scatter corrections of the annihilation 
photons, defining the LOR, into account~\cite{Poenisch03,Ollinger96} (while scattering effects
have been correctly simulated).

\section{Simulation and reconstruction setup}

 \begin{figure}
    \centerline{\includegraphics[width=0.8\textwidth]{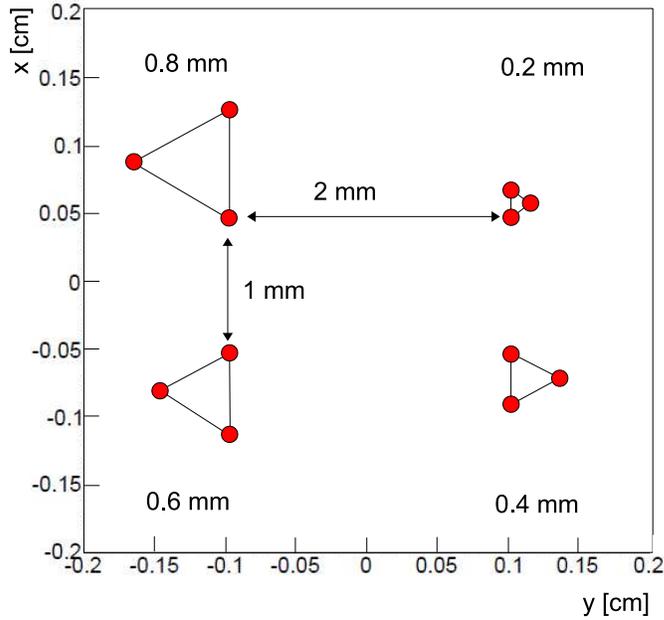}}
  \caption{Sketch of the simulated source geometry, representing a quasi-Derenzo
           phantom~\cite{Derenzo81}, consisting of twelve $^{22}$Na point sources with
           100 kBq activity each.} 
\label{fig:Der-Sketch}
\end{figure}

In order to characterize the spatial resolution of a PET scanner, a Derenzo phantom is
commonly used~\cite{Derenzo81}.
We simulated a quasi-Derenzo phantom (see Fig.~\ref{fig:Der-Sketch}
for a sketch of the source geometry),
consisting of twelve $^{22}$Na point sources with 100 kBq activity each.
Point sources were chosen for this exploratory study, while in a later
stage it is foreseen to extend this to the realistic scenario of an extended
source. Four equilateral
triangles are arranged in four sections, containing 3 point sources each, with
a separation of 0.2 mm, 0.4 mm, 0.6 mm and 0.8 mm, respectively.
This source arrangement was placed inside a water sphere of 6 cm diameter to imitate a
medical or biological sample and to take into account photon scattering effects
in the sample during the source reconstruction. Each of the four Compton camera 
modules consists of a LaBr$_{3}$ scintillator crystal (50 x 50 x 30 mm$^3$),
read out by a 2D-segmented photomultiplier with 64 pixels (6 x 6 mm$^{2}$ each).
An energy threshold of 5 keV (i.e. exceeding the electronic noise level) and an energy
resolution varying from $\Delta$E/E = 4.7\% at
500 keV to 3.5\% at 1 MeV were used. Furthermore,
 a double-sided silicon strip detector with 128 strips on each side, an active area of
50 x 50 mm$^{2}$ and a thickness of 2 mm was used as scatterer. The pitch size of 390 $\mu$m
correspondingly leads to a width of the LOR of 390 $\mu$m (FWHM).
An energy resolution
of 10 keV (FWHM) and a detection threshold of 10 keV in the Monte Carlo simulation
(due to the assumed electronic noise level of the DSSSD) was chosen. 
For the detector system a time resolution of 1~ns was (conservatively) assumed. The
rise time of signals from the LaBr$_{3}$ scintillator is about 8~ns, resulting in a
time resolution of 560~ps for our 50x50x30 mm$^3$ crystal.
With a typical signal rise time of around 2~ns for the silicon strip detector,
the time resolution of the combined detector system can be expected to be even
faster than 1~ns.
In order to test the feasibility of the $\gamma$-PET technique, Monte-Carlo simulations and
image reconstruction have been performed
using the 'Medium Energy Gamma-Ray Astronomy' library MEGAlib~\cite{Zoglauer06}.
MEGAlib is a software
framework designed to simulate and analyze data from Compton cameras. The library consists
of a Monte-Carlo simulation package, which utilizes the ROOT and Geant4 (9.4) software
library, an event reconstruction and an image reconstruction section based on a list-mode
maximum likelihood expectation maximization algorithm (LM-ML-EM). This algorithm is an
iterative method to reconstruct the most probable source distribution.
For the requirements of the $\gamma$-PET technique, we modified MEGAlib to realize an event
reconstruction from the intersection between the Compton cone and the LOR.
Subsequently, after successful event reconstruction, this information serves as starting
 point for an iterative image reconstruction of the $\gamma$-source positions.

\section{Results}
 \label{sec:results}

In the following we address the potential of the proposed $\gamma$-PET imaging
technique by separately discussing the main issues of spatial resolution, detection
efficiency and reconstruction sensitivity for different arrangements of the detection
system and imaged isotope.

\subsection{Spatial source reconstruction resolution}
 \label{subsec:resolution}

\begin{figure}
  \centerline{\includegraphics[width=0.8\textwidth]{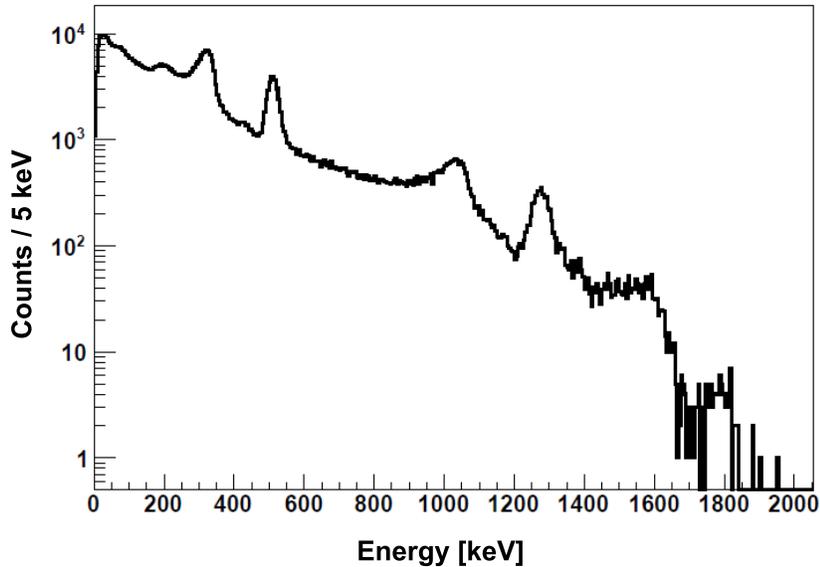}}
  \caption{$\gamma$-ray energy spectrum emitted from the twelve $^{22}$Na point sources,
           as detected in one of the 4 Compton camera modules placed outside a
           water sphere of 6 cm diameter. The point sources were arranged in
           the geometry of the Geant4 Monte-Carlo simulation as indicated in
           Fig.~2.} 
 \label{fig:Espec-detected}
\end{figure}

The $\gamma$-ray energy spectrum, as emitted from the twelve $^{22}$Na point sources,
and detected by one Compton camera module (including the detector resolution), placed
outside a water sphere of 6 cm diameter,
is shown in Fig.~\ref{fig:Espec-detected}. The spectrum was obtained from a
Monte-Carlo simulation using Geant4 (9.4).

\begin{figure}
      \centerline{\includegraphics[width=0.9\textwidth]{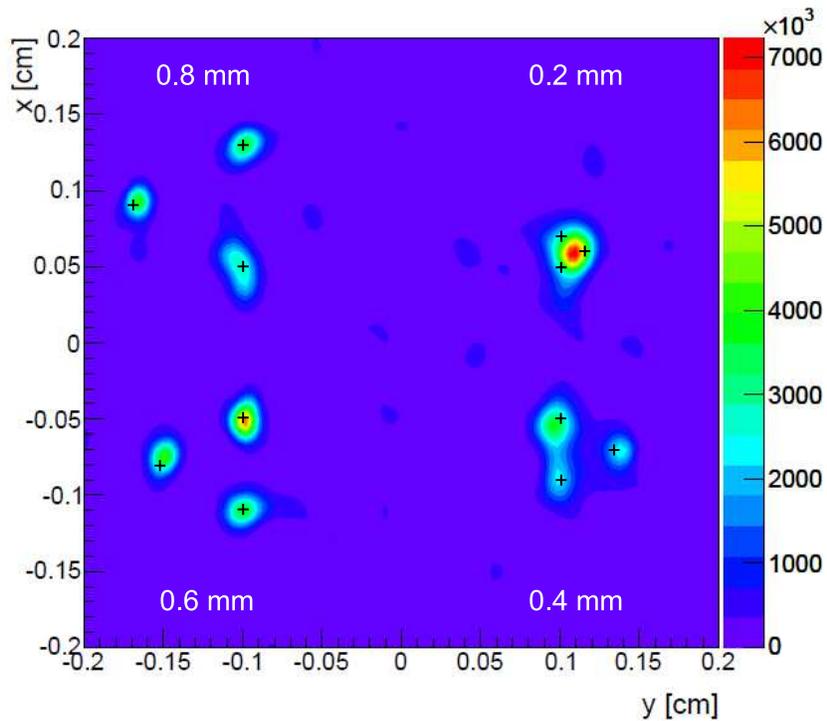}}
   \caption{Image of the reconstructed $\gamma$-source geometry of the quasi-Derenzo phantom
            introduced in Fig.~2, using the $\gamma$-PET technique
            after 100 iterations (depending on start parameters) using a
            maximum-likelihood algorithm. 
             The black crosses indicate the original source positions.}
   \label{fig:GPET-Derenzo}
\end{figure}

\begin{figure}
  \centerline{\includegraphics[width=0.8\textwidth]{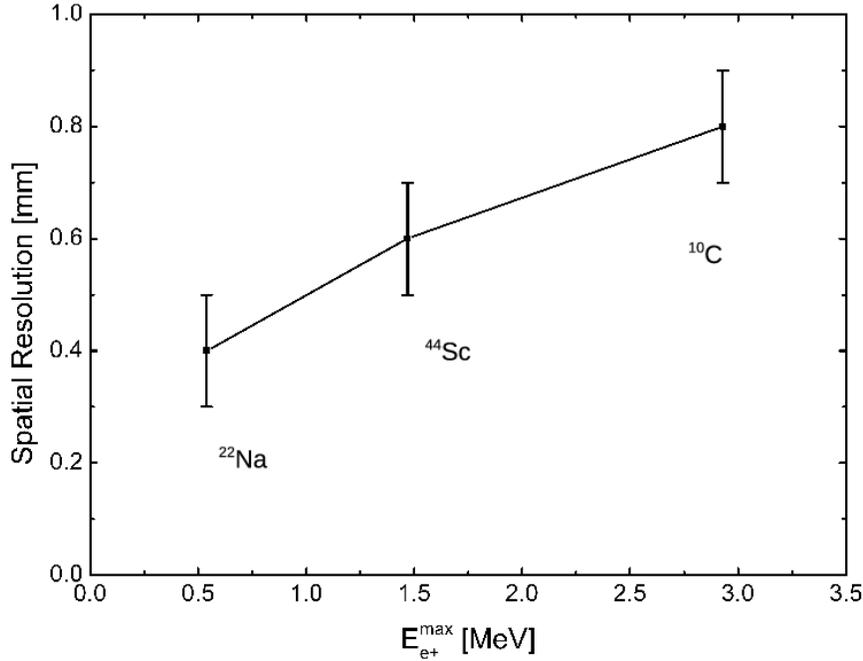}}
  \caption{Correlation between the spatial resolution (as estimated from an image
           reconstruction using the quasi-Derenzo phantom described before) and the
           $\beta$ end-point energy $E_{e^+}^{max}$ for the isotopes $^{22}$Na, $^{44}$Sc
           and $^{10}$C.}
    \label{fig:sRes_vs_Isotope}
\end{figure}

The decrease of of the $\gamma$-ray yield with energy is due to incompletely absorbed
$\gamma$ rays.
Clearly visible are the 511 keV positron annihilation line, as well
as the 1275 keV line from the $\gamma$ ray of the $\beta^+$ decay of $^{22}$Na.
The strong contribution at 340 keV comes from electrons due to Compton backscattering of 511 keV
while the peak at 1062 keV originates from Compton-backscattered 1275 keV photons.
The trigger condition in these simulations required three hits in three of the DSSSD modules
and one hit in one of the the LaBr$_{3}$ absorbers.
The line at 1786 keV is due to pileup between the 1275 keV transition and one of the
511 keV annihilation photons.
Based on the geometrical arrangement of $^{22}$Na sources (Fig.~\ref{fig:Der-Sketch}) and
detector modules (Fig.~\ref{fig:principle}a),
the underlying data of the detected $\gamma$-ray energy spectrum (Fig.~\ref{fig:Espec-detected})
are first used for a kinematical event reconstruction. The event reconstruction identifies
Compton events in an energy window of 1275 $\pm$ 50 keV, corresponding to the $\gamma$-ray
energy from the $^{22}$Na decay, also identifying simultaneous hits above the detection
threshold in the DSSSD for reconstruction of the LOR.
In Fig.~\ref{fig:GPET-Derenzo}, the resulting image of the reconstructed $\gamma$-source
geometry is shown, as obtained from exploiting the $\gamma$-PET technique. It was possible
to clearly resolve the two
largest triangles with spacings of 0.8 mm and 0.6 mm, respectively. The triangle with 0.4 mm
spacing still could be resolved sufficiently well, while the 0.2 mm spaced triangle could not
be resolved at all. The black crosses indicate the original source positions in the
simulation.
Due to the $\gamma$-PET technique, the imaging sensitivity for positron annihilation
significantly displaced from the initial decay spot via thermalization and diffusion
is strongly suppressed, and only positron annihilation photons emitted in
spatial and temporal coincidence with the third (prompt) $\gamma$ are included for image
reconstruction.
While the acollinearity of annihilation photons in our close detector geometry
(distance 50 mm to the source) contributes only about 0.3 mm to the position uncertainty,
Fig.~\ref{fig:sRes_vs_Isotope} shows the
correlation between the spatial resolution (as estimated via the above described
quasi-Derenzo phantom) and the $\beta$ end-point energy $E_{e^+}^{max}$ for the three
isotopes $^{22}$Na, $^{44}$Sc and $^{10}$C.
Isotopes with $E_{e^+}^{max} < 4$~MeV  are promising candidates for
sub-millimeter imaging in our geometrical detector arrangement.

\subsection{Efficiency considerations}
 \label{subsec:efficiency}

After having shown that the $\gamma$-PET technique allows for sub-millimeter spatial
resolution in the position reconstruction of the underlying radio-tracer independent of
its $\beta^+$ energy,
we address a further major advantage of this method, which is the acquisition time
reduction achievable with the highly sensitive triple-$\gamma$ coincidence measurement.\\
During the analysis of the 511 keV annihilation photons, no energy conditions have been
applied. The simulations showed that there is no necessity for a stricter event definition
than requiring an energy deposit (within a coincidence time window of 1 ns) above a
threshold of 10 keV in three DSSSD detectors (two of them diametral) and a
(completely absorbed) 1275 keV (prompt gamma from $^{22}$Na decay) signal from the
summed signal of the third scatterer and
its scintillator to generate the LOR and still reach sub-millimeter spatial resolution.
This can mainly be attributed to the capability of the event reconstruction algorithm
to provide a reliable reconstruction even on the basis of an incomplete photon energy
absorption.
Additional energy conditions would discard this latter class of events
(containing 5.8$\cdot$ 10$^{-2}$ (3.4$\cdot$~10$^{-3}$) of all events for
one (two) incompletely absorbed 511 keV photon(s)) and unnecessarily lead to a drastic
reduction of the reconstruction efficiency (see below). 
Here, 'reconstruction efficiency' denotes the ratio of the number of identified 
intersections between the LOR of the positron annihilation photons and the Compton cone 
derived from the detection of the third prompt photon and the number of initial decay events.
It comprises contributions from the photon detection efficiency as well as from the image
reconstruction sensitivity.
Moreover, the narrow timing
coincidence of 1~ns significantly helps to remove random background.
One individual Compton camera module simulated here
provides an event reconstruction efficiency of 3.3$\cdot$10$^{-5}$. Thus the geometry
studied with 4 camera modules exhibits an overall reconstruction efficiency for the
Compton cone of 1.3$\cdot$10$^{-4}$.
Moreover, a 5 (8) times thicker scatterer
(or a stack of 5 (8) scatter detectors with a summed thickness of 10 (16)~mm)
per Compton camera module would increase the reconstruction efficiency of
the Compton cones by an additional factor of 4.4 (5.8) to a value of
5.7 (7.5)~$\cdot$10$^{-4}$. Similar scatter detector thicknesses have been
favoured in \cite{richard2012}.
In the case of thicker scatterer, where no depth information of the scattering point
is measured, the spatial resolution increases from 0.4 mm (for 2 mm scatterer) to 0.6 mm
(for 16 mm scatterer).
This efficiency could be further increased by replacing our prototype geometry with a
pyramidal arrangement of a scatter detector and a larger absorber, covering the opening
angle of the cone seen from the photon source at the top of the pyramid.
For our detector geometry, this would require an absorber with an area of 114 x 114 mm$^2$,
about 5 times larger than the one used in our study. \\
Finally, when extending the Compton camera to a $\gamma$-PET device, the temporal and spatial
coincidence with the annihilation LOR has to be considered.
The simulated triple-coincidence detection efficiency for the $\gamma$-PET
technique (assuming the detector setup of Fig.~\ref{fig:principle}a))
amounts to 7.0~$\cdot 10^{-8}$ reconstructed intersections per $^{22}$Na decay
between the LOR of the annihilation photons and the Compton cone of the third photon.
This reduction of the above given Compton camera efficiency is on the one hand
due to the solid angle acceptance of the scatter detectors entering the LOR
reconstruction, in our scenario resulting in a geometrical coincidence probability of 0.026.
Moreover, a loss of those events has to be considered, where due to the diffusion of
the positron before its annihilation no intersection between its reconstructed LOR and
the Compton cone of the third photon, i.e. a spatial and temporal coincidence, could be found.
This fraction amounts to ca. 91.8$\%$, in total resulting in the above given overall
$\gamma$-PET reconstruction efficiency of 7.0~$\cdot$~10$^{-8}$
for $^{22}$Na.
This value would be further reduced to a prohibitively low value of
1.1~$\cdot$ 10$^{-10}$, if in addition to the condition set to the energy deposition
of 1275 keV in the scatterer and absorber from the third photon also conditions on the
energy of the two diametral 511 keV annihilation quanta (besides
their temporal coincidence within 1 ns) would have been required.
In a pyramidal arrangement of a scatter detector and a larger absorber covering the
full opening angle of the scatterer as seen from the emission point, such an
additional energy condition would not reduce the efficiency, but would reduce
random coincidences.
For an optimized detector system consisting of 4 Compton camera
modules, each with 8 x 2~mm thick scatter detectors and geometrically matched sizes
of the absorber crystals, a triple reconstruction efficiency of 9.7 $\cdot$ 10$^{-5}$
(for $^{22}$Na) can be expected.
The efficiencies of other isotopes are different, due to the individual $\gamma$ energies
$E_{\gamma}$, the different positron endpoint energies $E_{e^+}^{max}$ and the branching
ratio of $\gamma$/$\beta^+$.\\
The simulated reconstruction efficiencies for various $\beta^+\gamma$ emitters are
listed in Tab.~\ref{tab:efficiencies}.
\begin{table}[h]
\begin{center}
\begin{tabular}{l|cccccccc}
Isotope                            & $^{22}$Na & $^{44}$Sc & $^{14}$O & $^{68}$Ga & $^{124}$I & $^{10}$C &$^{76}$Br&$^{82}$Rb \\
\hline
$\epsilon_{\rm rec,1}$ [10$^{-8}$]  & 7.0       & 6.0      & 1.6      & 0.093     & 20        &  6.8    & 47     & 2.6      \\
\hline
$\epsilon_{\rm rec,2}$ [10$^{-5}$]  & 9.7       & 8.0      & 5.1      & 0.13      & 22        &  2.8    & 89     & 3.3      \\
\hline
\end{tabular}
  \caption{Simulated reconstruction efficiencies for various $\beta^+-\gamma$ emitters.
           In case of $^{124}$I and $^{76}$Br, which emit several prompt $\gamma$ rays with
           considerable branching ratio, the individual efficiencies have been summed up
          ($^{124}$I: 0.60 MeV, 0.72 MeV, 1.5 MeV and 1.7 MeV.
           $^{76}$Br: 0.56 MeV, 0.66 MeV, 1.13 MeV, 1.22 MeV and 1.85 MeV).
           The upper line corresponds to the detector system shown in
           Fig.~1a), while the lower line represents the reconstruction
           efficiency for an optimized system with thicker scatter and larger absorber
           detectors (see text).}
  \label{tab:efficiencies}
\end{center}
\end{table}

\subsection{Sensitivity considerations}
 \label{subsec:dose}

A minimum of 40 reconstructed intersections between the LOR and the
reconstructed emission direction of the third photon
is sufficient for a
reliable image reconstruction of a point source with a ratio of true-to-false reconstructed
events allowing for a correct reconstruction of the (sub-millimeter)
point-source position without fragmentation of the source image.
Choosing a typical injected activity of 400 MBq
and taking into account our intersection reconstruction efficiency $\epsilon_{\rm rec,2}$ for
$^{76}Br$, 40 intersections can be identified after an examination time
of about 140 seconds, which is sufficient for a reliable sub-millimeter
image reconstruction of a point source contained within a voxel volume of about 1~mm$^3$
('high-resolution mode').
In a standard PET Iterative Reconstruction analysis (without exploiting time-of-flight
information, i.e. ordered subset expectation maximization (OSEM)~\cite{Hudson94}),
about 6000 true coincidences acquired with a Siemens Biograph mCT PET scanner~\cite{siemens-mCT}
are necessary to localize a $^{22}$Na point source in the center of the scanner field-of-view
using the smallest voxel volume of 2 x 2 x 3 mm$^3$.
In order to compare the performance of the $\gamma$-PET technique with conventional
PET, we determined the imaging sensitivity of the method based on a comparable width
of the LOR of ca. 2~mm (in contrast to the previously used value for $^{22}$Na
of 0.4~mm). In such a case also about 40 reconstructed intersections per voxel
(derived without iterative reconstruction procedure) lead to a reliable localization
of the $^{22}$Na point source.
This sensitivity can be evaluated and compared to standard PET by quantifying the
examination time required to localize a point source as described above.
The minimum number of $\beta^+$-emitter decays per voxel $N_{\rm decay}$, required
for localizing a point source, relates to the imaging sensitivity, expressed by the
minimum number of reconstructed intersections $N_{\rm inter}$ and the corresponding
reconstruction efficiency $\epsilon_{\rm rec}$ as well as to the activity concentration
$C(i)$ of each voxel $i$, the examination time $\Delta t$ and the voxel
volume $V_{\rm vox}$ according to
\begin{equation}
 N_{\rm decay}(i) = N_{\rm inter}(i) / \epsilon_{\rm rec}
                 = C(i) \cdot \Delta t \cdot V_{\rm vox}(i) .
\end{equation}
For a given activity concentration and tumor size the required examination
time for localizing a point source only depends on the imaging sensitivity
$N_{\rm inter}$ and the reconstruction efficiency $\epsilon_{\rm rec}$.
While the $\gamma$-PET method described here features a clear advantage in terms
of sensitivity expressed by $N_{\rm inter}$ compared to standard PET, it falls behind
when comparing the corresponding efficiencies $\epsilon_{\rm rec}$, where values of about
0.1 are reported for small-animal PET scanners~\cite{Visser09}, while similar values
are found for whole-body scanners~\cite{Eriksson07}.\\
Tab.~\ref{tab:Typ-Isotope-Dose} lists in the last column the examination times
 $\tau^{exam}$ required for the localization of point sources of selected
$\gamma$-PET radioisotopes, where information of tracer and associated equivalent dose
coefficients (h$_T$) were available. $\tau^{exam}$ reflects the imaging sensitivity of
the $\gamma$-PET method based on the minimum requirement of 40 reconstructed intersections
and the isotope-specific reconstruction efficiencies $\epsilon_{\rm rec,2}$
(see Tab.~\ref{tab:efficiencies}).
Here we assume a typical value for the activity concentration accumulated in a
tumor of
25 kBq/ml, corresponding to PET examination of a patient with a body weight of
80 kg with an injected dose of 400~MBq and an average SUV (standard uptake value) of 5.
Column 7 shows the corresponding effective dose values (calculated with
h$_T$ values derived for adult males).
\begin{table}[h]
\begin{center}
\begin{tabular}{llccccccc}
Isotope&t$_{1/2}$&$\gamma$/$\beta^+$&Tracer&h$_T$ & Ref.& E$^{\rm exam}$&$\tau^{\rm exam}$&\\
       &  [min]    &         [$\%$]     &        & [$\mu$Sv/MBq]& &    [mSv] & [s] \\
\hline	  		    		   				
$^{14}$O   &  1.2 &  99 & Water    & 0.88 & \cite{Sajjad02}      & 0.33       & 2420   \\
$^{124}$I  & 6013 &  90 & MIBG     & 250  & \cite{Lee10}         & 92         & 558  \\
$^{76}$Br  & 16.2 & 100 & MAb-38S1 &  410 & \cite{Loevqvist99}   & 150        & 138   \\
$^{82}$Rb  &  1.3 &  13 & Chloride & 1.28 & \cite{Senthamizhchelvan11} & 0.46 & 3740 \\
$^{44}$Sc  & 236&     &  DOTATOC       & n.a.   &  \cite{Pruszynski10} &      & 1100 \\
\end{tabular}
\caption{$\gamma$-PET examination time $\tau^{exam}$ for the localization of
         a point source of selected $\beta^+-\gamma$-decaying radioisotopes, assuming an
         injected activity of 400 MBq and a tumor activity concentration of 25 kBq/ml.
         $\tau^{exam}$ reflects the imaging sensitivity of the $\gamma$-PET method
         based on the minimum requirement of 40 reconstructed intersections and the
         isotope-specific reconstruction efficiencies $\epsilon_{\rm rec,2}$.
         Column 7 shows the corresponding effective dose E$^{exam}$, based on the
         associated equivalent dose coefficients h$_T$ (for adult males).
         The first columns list decay properties and suitable tracers of the respective
         isotopes.}
\label{tab:Typ-Isotope-Dose}
\end{center}
\end{table}		
The rather wide spread of $\tau^{\rm exam}$ values, reflecting the corresponding 
isotope-specific spread of the reconstruction efficiency $\epsilon_{\rm rec}$,
exhibits no clear correlation with any of the isotopic properties like the
$\beta^+$ endpoint energy or the energy of the third prompt photon. Thus it may
rather represent a correlated interplay between different factors that makes it
difficult to predict the performance of a specific radioisotope when applying the
$\gamma$-PET technique.
In case of small-animal PET, where even higher effective doses are injected, real-time
imaging of the metabolism or organ motion, or a study of biological washout
diffusion processes in animal experiments with implanted
radioisotopes~\cite{mizuno03}, appears feasible with the $\gamma$-PET technique.

\section{Conclusion and Outlook}

Most medical radioisotopes typically give rise to a lower spatial resolution for PET
imaging,
compared to the most widely used $^{18}$F, due to their higher $\beta^+$ decay energies,
resulting in a larger positron diffusion range.
We investigated the '$\gamma$-PET' imaging technique, taking advantage of detecting the
additionally emitted $\gamma$ ray in coincidence with the $\beta^+$ annihilation photons.
The triple-coincidence measurement allows to reduce the image-blurring effect of the
diffusion range of the positron prior to its annihilation, which increasingly gains importance
when comparing radioisotopes with higher $\beta^+$ endpoint-energies compared
to the 634 keV for $^{18}$F.
Since in analogy to PET, Compton scattering of the annihilation radiation within the 
patient limits the imaging performance,
a further improvement of the $\gamma$-PET technique would have to take attenuation and scatter
corrections of the annihilation photons, defining the LOR, into
account~\cite{Poenisch03},~\cite{Ollinger96}.
Simulations showed that it is possible to reach sub-millimeter spatial resolution in case
of a small-animal imaging scenario, i.e. a small distance between the source and the
detector, where the limiting influence of the acollinearity can be neglected.
Even in case of high-energy positron emitting isotopes like $^{76}$Br
($E_{e^+}^{max}$ = 3.38 MeV) or $^{10}$C ($E_{e^+}^{max}$ = 2.93 MeV),
the image reconstruction will again result in sub-millimeter spatial resolution.
Moreover, being left only with the limiting effect of the acollinearity for whole-body
PET scanning, most of the radioisotopes listed in Tab.~\ref{tab:Isotope-Range} still
allow to reach sub-millimeter spatial resolution also for a clinical scenario.\\
One Compton camera module as described in Fig.~\ref{fig:principle}, having a rather
limited field-of-view, exhibits a Compton-cone reconstruction efficiency of
3.3$\cdot$10$^{-5}$. However,
the $\gamma$-PET technique requires at least 3 camera modules. We simulated
different detector arrangements for a $\gamma$-PET prototype consisting of 4 Compton
camera modules, resulting for an optimized setup with thick scatter detectors and large
absorbing crystals - matched to the solid angle covered by the scatterer - in
an isotope-dependent triple-photon intersection reconstruction efficiency
between 1.3~$\cdot 10^{-6}$ ($^{68}$Ga) and 8.9~$\cdot 10^{-4}$ ($^{76}$Br).\\
Particularly attractive is the highly sensitive image reconstruction capability
provided by the $\gamma$-PET technique, found superior to a conventional PET
scanners. Presently the full potential of this advantage cannot be exploited,
due to the much reduced reconstruction efficiency compared to conventional full-body
or small-animal PET scanners. Thus research and development efforts should be directed 
towards optimizing the efficiency achievable with $\gamma$-PET. 
Moreover, the present study was limited
to the performance involving a point source in a scattering medium, while further work 
will also address
the characterization of the method with respect to extended photon sources.\\
Finally, the Compton camera described here could also turn out to be beneficial in a
therapeutic hadron beam irradiation, where $\beta^+(\gamma)$ emitters ($^{10}$C, $^{14}$O)
are generated via, e.g., the carbon beam.
Especially the projectile (fragment) $^{10}$C with its short half-life of 19.3~s
and the quasi-simultaneous emission of a third photon from an excited state qualifies
as an online marker isotope during hadron therapy. Its spatial distribution within
the patient could be tomographically reconstructed, either from a (quasi-realtime)
PET analysis (i.e. direct reconstruction using TOF-PET) or using the hybrid
$\gamma$-PET technique to achieve an improved spatial resolution together with
an enhanced sensitivity, i.e. reduced requirements to the signal strength.
The presented Compton camera (eventually upgraded by a thicker scatterer), could
provide a versatile setup to assist with targeting one of the crucial issues of
hadron therapy, which is ion beam range verification, either by detecting
prompt $\gamma$ radiation during the irradiation~\cite{Min06} or (delayed)
short-lived $\beta^+$-decaying reaction products
(PET- or $\gamma$-PET operation) in between the irradiation cycles.
Exploiting the perspectives of the $\gamma$-PET technique may thus allow to turn
the present disadvantages of $\beta^+\gamma$-emitting PET isotopes into a benefit
in resolution or sensitivity.

\acknowledgments

We are indebted to Christopher Kurz from the Heidelberg Radiooncology Department
for his help and committment in deriving the statistics limits for conventional
PET imaging. Also, fruitful discussions with Dr. Guido B\"oning from the LMU
Nuclear Medicine Department are gratefully acknowledged.
This work was supported by the DFG Cluster of Excellence MAP (Munich-Centre for Advanced
photonics).

\end{document}